\documentclass[runningheads,a4paper]{llncs}

\usepackage{amssymb}
\usepackage{graphicx}
\usepackage{mathptmx}       % selects Times Roman as basic font
\usepackage{helvet}         % selects Helvetica as sans-serif font
\usepackage{courier}        % selects Courier as typewriter font
\usepackage{type1cm}        % activate if the above 3 fonts are
\usepackage{makeidx}         % allows index generation
\usepackage{graphicx}        % standard LaTeX graphics tool
\usepackage{multicol}        % used for the two-column index
\usepackage[bottom]{footmisc}% places footnotes at page bottom
\usepackage[table]{xcolor} %multi-column tables
\usepackage{epsfig, tkz-berge,tikz,pgf,tkz-graph}
\usepackage{color, hyperref}
\usepackage{array}
\usepackage{hyperref} 
\usepackage[mathscr]{euscript}
\usepackage{graphicx,color,ulem,dsfont}
\usepackage{lineno}
\usepackage{epsfig, tkz-berge,tikz,pgf,tkz-graph}
\usepackage{float}
\usepackage{algorithm}
\usepackage{color}
\usepackage[noend]{algpseudocode}
\usepackage{centernot}
\usepackage{caption}
\usepackage{subcaption}
\usepackage{multicol}
\usepackage{enumerate}
\usepackage{relsize}
\usepackage{comment}
\usepackage[utf8]{inputenc}
\usepackage{natbib}
\usepackage{float}
\usepackage{xspace}

\usepackage{url}
  \let\oldurl\url
\usepackage{hyperref}
  
  \let\url\oldurl

\usepackage{amssymb,amsbsy,amsfonts,latexsym,verbatim}
\usepackage{caption}

\usepackage{url}

\newcommand{\keywords}[1]{\par\addvspace\baselineskip
\noindent\keywordname\enspace\ignorespaces#1}

\begin{document}

\mainmatter  % start of an individual contribution

% first the title is needed
\title{Sampling a Network to Find Nodes of Interest}

% a short form should be given in case it is too long for the running head
\titlerunning{Sampling a Network to Find Nodes of Interest}

% the name(s) of the author(s) follow(s) next
%https://www.sharelatex.com/project/579b90be5b2bcf846e625587
% NB: Chinese authors should write their first names(s) in front of
% their surnames. This ensures that the names appear correctly in
% the running heads and the author index.
%
\author{Pivithuru Wijegunawardana$^1$ \and Vatsal Ojha$^2$ \and Ralucca Gera$^3$% 
% I listed the authors in alphabetical order, feel free to re-arrange if you wish
\thanks{Ralucca Gera would like to thank the DoD for partially sponsoring the  current
research.}%
\and Sucheta Soundarajan$^1$}
\authorrunning{Wijegunawardana,  Ojha, Gera, Soundarajan,}
% (feature abused for this document to repeat the title also on left hand pages)

% the affiliations are given next; don't give your e-mail address
% unless you accept that it will be published
\institute{$^1$ Syracuse University, Department of Electrical Engineering \& Computer Science,\\ \email{ppwijegu,susounda@syr.edu}\\$^2$ Dougherty Valley High School, San Ramon, California, \email{vatsalojha@gmail.com}\\
$^3$Naval Postgraduate School, Department of Applied Mathematics, Monterey, CA,
\email{RGera@nps.edu}
%\mailsa\\
%\mailsb\\
%\mailsc\\
%\url{http://www.springer.com/lncs}
}

%
% NB: a more complex sample for affiliations and the mapping to the
% corresponding authors can be found in the file "llncs.dem"
% (search for the string "\mainmatter" where a contribution starts).
% "llncs.dem" accompanies the document class "llncs.cls".
%

\toctitle{Sampling a Network to Find Nodes of Interest}
%\tocauthor{Authors' Instructions}

\maketitle

\begin{abstract}
The focus of the current research is to identify people of interest in social networks.  We are especially interested in studying dark networks, which represent illegal or covert activity. In such networks, people are unlikely to disclose accurate information when queried.  We present {\sc RedLearn}, an algorithm for sampling dark networks with the goal of identifying as many nodes of interest as possible.  We consider two realistic lying scenarios, which describe how individuals in a dark network may attempt to conceal their connections.  We test and present our results on several real-world multilayered networks, and show that {\sc RedLearn} achieves up to a 340\% improvement over the next best strategy.
\end{abstract}
%We need to stay within 12 pages

%%%%%%%%%%%%%%%%%%%%%%%%%%%%%%%%%
\section{Introduction and Motivation}  \label{section:introduction}

The advent of big data in today's complex environment requires decision makers to act in an overwhelmingly rich environment (or network) based on partial information of that network. Using an observed portion of some network, it is often desirable to make inferences about which unobserved areas of the network to explore next. Often, it is of particular interest to locate ``people of interest" (POI) residing in these networks.  These individuals may be trying to conceal themselves, or may be protected by others.

Our work was motivated by study of terrorist networks.  These networks typically include several types of relationships connecting people, producing multilayered networks where each layer is defined by a different relationship.  For example, one of the networks contains data depicting relationships among terrorists in the 'Noordin Top' Network in Indonesia,  where relationships are organizations these terrorists belong to, the schools or trainings they went to, kinship, recruiting and so on.  The data was collected by  Everton et. al~\cite{noordin} and compiled into a network by Gera et al. in~\cite{DarkNetworks}.    

It is generally desired to find those POIs that possess a certain attribute, such as people who attended a specific planning meeting or who were involved in organizing a particular attack. For our specific work, in the insurgent network Noordin Top (see Section~\ref{subsection:Data}) all the people that were communicating using a certain medium were tagged as being POI, producing a relatively small number of POI.  We assume that we begin with knowledge of one POI in the network, with the rest of the network unobserved (both in terms of topology as well as node attributes).  Our goal is to sample the network in such a way that we observe as many POIs as possible.  To our knowledge, we are the first to consider the problem of sampling with the goal of identifying nodes of interest in a setting with misinformation.

We present {\sc RedLearn}, a novel, learning-based algorithm for sampling networks with the goal of finding as many POIs as possible.  We show that in cases where the POIs exhibit homophily (i.e., are likely to be connected to other POIs), a simple strategy of choosing the node with the most POI neighbors works well.  However, in the more realistic scenario where POIs hide their connections with other POIs, {\sc RedLearn} shows outstanding performance, beating the next best strategy by up to 340\%.

%%%%%%%%%%%%%%%%%%%%%%%%%%%%%%%%%
\section{Problem Definition}  \label{section:problem}
%%%what: Talk about monitors, colors, notations etc. and introduce problem formally
We refer to nodes representing POIs as `red' nodes, and other nodes as `blue', giving us a purple network.  We assume that there is an unobserved, underlying graph $G = (V, E)$, in which each node $v \in V$ has color $C_v \in \{red, blue\}$.  We begin with having knowledge of only one red node in $G$.  

To increase our observation of the network, we place \textit{monitors} on nodes.  A monitor tells us (1) the true color of the node being placed on, (2) the true neighbors of that node, and (3) the colors of the node's neighbors, possibly with inaccuracies.  For example, placing a monitor on a suspected terrorist could represent determining whether that person is actually a terrorist, determining who his or her e-mail or phone contacts are, and questioning the individual about whether those neighbors are themselves terrorists.  Naturally, some individuals may lie about the colors of those neighbors.\footnote{We consider two realistic 'lying scenarios'; these are described in Section~\ref{sec:lying-strategy}.}  

We assume that we are given a budget of $b$ monitors, and can place those monitors on any node that has been observed.  In the first step, we must place a monitor on the initially observed node, because no other nodes were observed.  In subsequent steps, we may place a monitor on any node that has been observed as a neighbor of a previously-monitored node.  

Our goal is to maximize the total number of red nodes observed.  

%%%%%%%%%%%%%%%%%%%%%%%%%%%%%%%%%
\section{Related Work}  \label{section:related-work}
%dark networks
Our work here is related to work on analyzing dark networks, a special type of social network~\cite{baker1993social}.   A dark network is network that is illegal and covert~\cite{raab2003dark}, whose members  are actively trying to conceal network information even at the expense of efficiency~\cite{baker1993social}, and the existing connections  are used  infrequently~\cite{raab2003dark}.  Because a dark network is deceptive by nature, we examine the lying methodologies along with the discovery methods in looking for the POI.

%sampling
There are a multitude of sampling techniques for network exploration, including random walks (\cite{asztalos2010network}, \cite{hughes1996random}, \cite{noh2004random}), biased random walks (\cite{fronczak2009biased}), or walks combined with reversible Markov Chains(\cite{aldous2002reversible}), Bayesian methods(\cite{friedman2003being}), or standard exhaustive search algorithms like depth-first or breadth-first searches, such as ~\cite{adamic2001search, biernacki1981snowball,  bliss2014estimation, davis,  leskovec2006sampling}.  However, these methods fail in using discovered knowledge, such as node attributes, effectively.  %However, not much has been done in the realm of the mulilayered networks as our current project proposes. 

Various researchers have considered the problem of sampling for specific goals, such as maximizing the number of nodes observed.  For example, Avrachenkov, et al. present an algorithm to sample the node with the highest estimated unobserved degree~\cite{Avrachenkov2014MEUD}.  Hanneke and Xing~\cite{Hanneke2009Sample}, and Maiya and Berger-Wolf~\cite{Maiya2013Crawling} examine online sampling for centrality measures.  Macskassy and Provost develop a guilt-by-association method to identify suspicious individuals in a partially-known network~\cite{macskassy2005suspicion}.

%a paragraph on sampling multilayered networks

%a paragraph on POI

%%%%%%%%%%%%%%%%%%%%%%%%%%%%%%%%%%%%%%%%%%%%%%%%%%%%%%%%%%%%%%%%%%%%%%%%%%%%%%%%%%%%%%%%%%%%%%%%%%
\section{Proposed Method: {\sc RedLearn}}
A monitor placement strategy is an incremental sampling strategy.  In each step, it uses current knowledge of the discovered graph, including the observed topology of the graph, known colors of nodes (observed by monitors placed directly on those nodes), and the stated colors of monitored nodes' neighbors (i.e., for each neighbor of a monitored node, whether the monitored node said that that neighbor was red or blue).  With this information, the strategy determines where to place the next monitor to discover the most number of red nodes.  This continues until the monitor budget is exhausted.

\textit{monitored node} is a node with a monitor placed on it.  A monitored node is \textit{known red (or blue)}.  Recall that a monitored node may lie about its neighbors' colors; thus the color of a node is not known with certainty until it is monitored. Let $N(v)$ refer to the set of neighbors of $v$.

%how
\subsection{Baseline Monitor Placement Strategies}
We now describe several natural monitor placement strategies.  We use these strategies as comparison algorithms in our experiments; additionally, some of these strategies are used in developing our learning strategy {\sc RedLearn}.

%The strategies we consider are:  list them all.
\textbf{Smart Random Sampling (SR):} In each step, the Smart Random Placement strategy places monitors on a random unmonitored node. Due to the inefficiency of using no information about the network for monitor placement, this is used as a lower bound placement strategy for the introduced methodologies. 

\textbf{Red Score (RS):}  The Red Score strategy is guided by the colors reported by neighbors of a node. If a node $v$ reports its neighbor $u$ as red, the score associated with node $u$ is increased by one, making it more suspicious. This strategy selects the node with highest red score to place the next monitor. For this method, the red score is highly impacted by the accuracy of information given by the neighboring node. Additionally, due to its use of both red and blue node information, this strategy uses the most amount of information as compared to the other baseline strategies.

\textbf{Most Red Say Red (MRSR):} The MRSR strategy places a monitor on the node with the greatest number of red neighbors who report it as a red node. It does not factor in blue node information and is dependent solely on the accuracy of the information given by neighboring red nodes. Blue nodes are essentially useless in this strategy, mimicking the reality when they might not know who the POIs are. This placement strategy would result in a red node with no red neighbors being impossible to discover except by chance.

\textbf{Most Red Neighbors (MRN):} The MRN placement strategy places a monitor on the node with the most known red neighbors. This strategy ignores what the monitored nodes say about their neighbors. This strategy would likely work best in a network with high homophily. Similar to the MRSR strategy, blue neighbors are unimportant in determining the likelihood of a given node being red.

\subsection{{\sc RedLearn}: A Learning Based Monitor Placement Strategy}\label{sec:lying-strategy}
%%Why we need a learning based model: Each score consider only the colors others say or network structure, we wanted to build a score that will use all information available

%%Explain classification model and classifier used (logistic regression)

%%Explain features and why we used each

When determining which node $v$ to place the next monitor on, for each candidate node $v$ (i.e., each observed but unmonitored node), the strategies above consider the colors of $v$'s neighbors and/or the color that each of $v$'s monitored neighbors reported. 

Intuitively, if the original network displays homophily, the probability of node $v$ being a red node is higher if it has many red neighbors. But if the network does not display homophily, using this measure can result in  poor performance.  Note that we expect many dark networks to exhibit low or even anti-homophily, as malicious actors are likely to go out of their way to conceal their connections.  

Additionally, some monitor placement strategies like red score, depend on information given by neighbors. The performance of such monitor placement criteria thus heavily depend on the policies that nodes follow when stating their neighbors' colors.

To overcome these dependencies, we propose {\sc RedLearn}, a learning based monitor placement strategy. Our goal is to successfully predict the probability of a node $v$ being red ($P(v=R)$) based on the observed network structure and what $v$'s neighbors say about $v$. We model this as a two class classification problem, but rather than looking at the assigned label (Red or Blue), we are more interested in finding $P(v=R)$.  Once these probabilities are determined, {\sc RedLearn} places the next monitor on the node with the highest such probability. 

\textbf{Features: } Table~\ref{table:features} describes the set of features used in our learning based monitor placement algorithm.  There are two types of features: ($a$) Network structure-based features ($1, 2, 3$), and ($b$) Neighbor answer-based features ($4, 5, 6, 7, 8$).
\vspace{-1.1cm}
 \begin{center}
    \begin{table}[!htbp]
    \caption{Classification features for {\sc RedLearn}. Consider a node $v$ with neighbors $N(v)$}
    \begin{tabular}{|c|c|c|}
        \hline
         & Feature &  Description \\ 
        \hline
        \hline
        (1)&Number of Red Neighbors  &  $\displaystyle |\{u \in N(v) | c_u=R\}|$ \\
        (2)&Number of Blue neighbors &  $ \displaystyle|\{u \in N(v) | c_u=B\}|$ \\
        (3)&Number of Red triangles if $v$ is red     &  $\displaystyle \bigg |\{u,w \in N(v) | u \in N(w) \cap w \in N(u) \cap c_u=c_w=R\}\bigg|$ \\
        (4)&Red score &  $\displaystyle |\{u \in N(v) | (u \; says \; R) \}|$  \\
        (5)&Number of Red neighbors saying red  &  $\displaystyle |\{u \in N(v) | (u \; says \; R) \cap c_u=R \}|$ \\
        (6)&Number of red neighbors saying blue  & $\displaystyle |\{u \in N(v) | (u \; says \; B) \cap c_u=R \}|$ \\
        (7)&Number of blue neighbors saying red  &  $\displaystyle |\{u \in N(v) | (u \; says \; R) \cap c_u=B \}|$ \\
        (8)&Number of blue neighbors saying blue  &  $\displaystyle |\{u \in N(v) | (u \; says \; B) \cap c_u=B \}|$ \\
        (9)&Inferred probability of being red  &  $ \displaystyle P^I(v=R)$ \\
        \hline
    \end{tabular}
        \label{table:features}
    \end{table}
    \end{center}    
\vspace{-.9cm}
 Network structure-based features are used to learn the patterns of connections between red nodes (e.g., homophily vs. anti-homophily). Neighbor answer-based features are intended to learn the relationship between what a node says about its neighbors' colors and the true colors of those neighbors.

\textbf{Inferred probability of being red: } 
%Estimating the underlying lying model (i.e., whether a monitored node lies or is honest about its neighbors' colors) helps to identify red nodes in the graph.
We formulate four different probabilities to measure the trustworthiness of colors given by differently colored nodes (i.e., whether a monitored node lies or is honest about its neighbors' colors). We use the information collected prior to placing a monitor on a node to calculate these probabilities.  Consider a node $v$ which was discovered through a monitor placed on node $u$. 
\begin{enumerate}
    \item $\displaystyle P(v=R|(u=R) \wedge (u \; Says \; R))= \frac{|\{(v=R) \cap (u=R) \cap (u\;says\;R)\}|}{|\{(u=R) \cap (u\;says\;R)\}|}$
     \item $\displaystyle P(v=R|(u=R) \wedge (u \; Says \; B))= \frac{|\{(v=R) \cap (u=R) \cap (u\;says\;B)\}|}{|\{(u=R) \cap (u\;says\;B\}|}$
      \item $\displaystyle P(v=R|(u=B) \wedge (u \; Says \; R))= \frac{|\{(v=R) \cap (u=B) \cap (u\;says\;R)\}|}{|\{(u=B) \cap (u\;says\;R)\}|}$
      \item $\displaystyle P(v=R|(u=B) \wedge (u \; Says \; B))= \frac{|\{(v=R) \cap (u=B) \cap (u\;says\;B)\}|}{|\{(u=B) \cap (u\;says\;B)\}|}$
\end{enumerate}

%These probabilities give general information about the probability of a node $v$ being a red node if a red neighbor $u$ says it is red, if a red neighbor $u$ says it is blue, if a blue neighbor $u$ says it is red, and a blue neighbor $u$ says it is blue.
Given a node $v$, we calculate the inferred probability, $P^I(v=R)$ using equation~\ref{eq:global_prob}. 

%%We run our monitor placement strategies starting from one red node and continue to probe network as we place monitors. We get to know the neighbors of a node, only if we place a monitor on that node. Because of this probing strategy most non-monitor nodes tend to be leaf nodes or have few neighbors. This results in having several nodes with the same score for the monitor placement strategy in consideration. Now this probability is used to break ties among nodes with similar scores for monitor placement strategies.

\begin{equation}\label{eq:global_prob}
\displaystyle P^I(v=R) \;  =  \; \frac{\sum_{u \in N(v)}{P(v=R| color(u) \wedge color(u \; says \; v))} }{|N(v)|}  
\end{equation}

\textbf{Training Data:} Suppose that we have placed $n$ monitors so far.  Then the training set consists of the true colors of the $n$ monitored nodes along with their respective feature values. To determine where to place the $(n+1)^{st}$ monitor, we train the learning model using this data.\\

\textbf{Classification Algorithm:} Our goal is finding $P(v=R)$ for each unmonitored node $v$. Because we are predicting a probability rather than a binary label, we propose using a logistic regression classifier. Furthermore, because the learning model must be updated frequently, this classifier gives an added advantage of faster training.\\

\textbf{Placing the Next Monitor (Prediction):} Given the placement of $n$ monitors and deciding to place  the $(n+1)^{st}$ monitor, {\sc RedLearn} considers all unmonitored nodes discovered so far. Next, it calculates feature vectors associated with these non monitor nodes, and applies the classifier to these feature vectors, giving the probability that each unmonitored node is red. {\sc RedLearn} selects the node with the highest probability for placing the next monitor.  Algorithm~\ref{algo:learning} summarizes {\sc RedLearn}.

\begin{algorithm}[!htbp]
\caption{Learning based monitor placement}\label{algo:learning}
\begin{algorithmic}
\Procedure{Learning}{\textit{start,budget}}
\State $\textit{G} \gets$ Graph
\State $\textit{G}$.add(start), $\textit{G}$.add($N(start)$) \Comment{Starting node and neighbors}

\While{\textit{budget$>$0}}
    \State $\textit{Monitors} \gets$ list of monitored nodes in \textit{G}
    \State  $\textit{TrainingData} \gets$ feature vectors for $\textit{Monitors}$
    \State Train classifier using $\textit{TrainingData}$
    \State $\textit{NotMonitors} \gets$ list of not yet monitored nodes in \textit{G}
    \For{$v$ $\in$ $NotMonitors$}
    \State  Get feature vector for $v$
    \State $\textit{P(v=R)} \gets$ predict $v$'s probability of Red using learning model
    \EndFor
    \State Choose node $v$ with maximum $P(v=R)$ from \textit{NotMonitors}
    \State $\textit{budget} \gets (budget-1)$
    \State Use $v$ as next monitor
    
\EndWhile
\EndProcedure
\end{algorithmic}
\end{algorithm}
%\vspace{-0.5cm}

% \begin{algorithm}[!htbp]
% \caption{Smart Random Placement}\label{SRP}
% \begin{algorithmic}
% \Procedure{Random Placement}{\textit{G}, \textit{Monitors}}
% \State $\textit{G} \gets$ Graph
% \State $\textit{NotMonitors} \gets$ list of not yet monitored nodes in \textit{G}
% \While{\textit{All red nodes not monitored}}
%     \State Choose random node $u$ from \textit{NotMonitors}
%     \State Remove $u$ from \textit{NotMonitors}
%     \State Use $u$ as next monitor
    
% \EndWhile
% \EndProcedure
% \end{algorithmic}
% \end{algorithm}

% \begin{algorithm}[!htbp]
% \caption{Red Score}\label{RS}
% \begin{algorithmic}
% \Procedure{Update Red Score}{\textit{G}, \textit{Monitors}}
% \State $\textit{G} \gets$ Graph
% \State $\textit{NotMonitors} \gets$ list of not yet monitored nodes in \textit{G}
% \State $S_u \gets$ Red Score of node $u$
% \State $C_u \gets$ Color of node $u$
% \State $N_u \gets$ Neighbor List of $u$
% \While{\textit{All red nodes not monitored}}
%     \State $u \gets$ Monitor
%     \State Remove $u$ from \textit{NotMonitors}
%     \For{$v$ in $N_u$}
%         \If{$u$ says $C_v$ is Blue}
%             \State Add value to $S_v$
%         \EndIf
%         \If{$u$ says $C_v$ is Red}
%             \State Subtract value to $S_v$
%         \EndIf
%     \EndFor
%     \State Choose node with lowest Red Score from \textit{NotMonitors} as next Monitor
% \EndWhile
% \EndProcedure
% \end{algorithmic}
% \end{algorithm}

\section{Experimental Set Up}

In section \ref{subsection:Data}, we give a description of our network datasets, and then consider them without homophily as described in Section~\ref{subsection:homophily}. In Section \ref{subsection:lying-strategy}, we introduce two lying scenarios to model the lying behavior of a node (i.e., whether it says its neighbors are red or blue). Finally, we describe our experimental setup.

\subsection{Datasets}\label{subsection:Data}

\textbf{Noordin Top Network:}  The first network studied is Noordin Top, a relatively small, but real network with $139$ nodes and $1042$ edges depicting several types of relationships between them (`Noordin Top' is the name of the leader of this network).\footnote{Obtained from \url{https://sites.google.com/site/sfeverton18/research/appendix-1}. %\cite{noordin}
} In this network, every node is a terrorist, and we are interested in identifying those individuals that communicate using a certain medium.  We consider 5 versions of this network.  In NoordinComs1, the POI (red nodes) are those communicating using a general computer medium; in NoordinComs2, the red nodes are those who communicate using print media; in NoordinComs3, the red nodes are those who communicate using support materials; in NoordinComs4, the red nodes are those who communicate using unknown media; in NoordinComs5, the red nodes are those who communicate using video (Table~\ref{table:NoordinOverview}). 
    
    \vspace{-1cm}
    \begin{center}
    \begin{table}[!htbp]
            \caption{Noordin Top network overview}
        \label{table:NoordinOverview}
    \begin{tabular}{|c||c|c|c|}
        \hline
        \hspace{.8cm}Network name \hspace{.8cm} & \hspace{.8cm} Red Node Count\hspace{.8cm}  & Degrees of Red Nodes \hspace{1cm}\\ 
        \hline
        \hline
        NoordinComs 1 &  9  & 8,12,20,21,33,38,50\\
        NoordinComs 2 &  5  & 8,21,38,38,50\\
        NoordinComs 3 &  9  & 11,12,21,38,39,50,52\\
        NoordinComs 4 &  18 & 17,21,24,27,31,38,40,45,52,53,58\\
        NoordinComs 5 &  11 & 0,9,21,33,38,41,50,52\\
        \hline
    \end{tabular}
    \end{table}
    \vspace{-.6cm}
    \end{center}

\textbf{PokeC Network:}  The PokeC network is part of a Slovenian on-line social network.\footnote{Obtained from \url{http://snap.stanford.edu/data/}.} The nodes in the network are users of the social network and edges depicts friendship relations. Each node has some number of associated user attributes (e.g., age, region, gender, interests, height etc.).  We use a sample of this network containing all nodes in the region "kosicky kraj, michalovce" and edges among them. This sampled network contains $26,220$ nodes and $241,600$ edges.

We assign node colors based on two different node attributes:
%to retrieve two networks to run our algorithm. The first attribute we selected was 
\textit{age} (a node with age in the range 28-32 is marked red, and blue otherwise, giving $1736$ red nodes) and \textit{height} (a user of height less than $160\;cm$ is marked red, giving $1668$ red nodes).

\subsection{Eliminating Homophily}\label{subsection:homophily}
In both networks that we consider, red nodes tend to be connected to each other. However, in a dark network where red nodes are actively trying to hide their presence, these nodes would conceal the existence of such connections (for example, instead of using their normal cell phone to make calls to other red nodes, a red node might use a burner phone for such calls).  To account for this, we consider versions of our datasets where all connections between red nodes are removed.  Note that this type of network presents a much more challenging setting, as one cannot simply rely on homophily to find red nodes.

\subsection{Lying Scenarios}\label{subsection:lying-strategy}
% why do we need to model lying behavior, factors that affect u lying about V, 3 lying strategies, how to get the prob(u lie about v) in three lying strategies
Recall that a monitor tells us the true color of the monitored node and the possibly incorrect colors of that node's neighbors.  Because we do not have data describing how terrorists lie to prevent detection of other terrorists, we must simulate this behavior.  

In formulating these lying scenarios, we assume the existence of a hierarchy among the nodes, where nodes are more likely to lie to protect those above them in the hierarchy.  We assume that the red nodes are fully aware of the hierarchy, but blue nodes may or may not be aware, depending on the scenario.

In both lying scenarios, we assume that nodes may lie not only about the color of red nodes (i.e., lie to protect POIs), but also about the color of blue nodes (i.e., framing innocent individuals as a distraction).  Without this assumption, the problem would be trivial, because anytime \textit{any} individual said a node was red, we would know with full certainty that that node is actually red.

Consider nodes $u$ and $v$, where $u,v \in Edges$. The probability that $u$ lies about $v$, $P(u \; lie \; v)$ depends on:
\begin{itemize}
    \item The color of $u$ ($C_u$) and color of $v$ ($C_v$).
%    Whether $u$ knows $v$'s true color depends on the color of $u$. If $u$ is a blue node, it might not know about red nodes. Further, whether $v$ needs to be lied about depends on its color as well. 
    \item The inherent honesty of $u$ ($H_u$), where higher $H$ values indicate that $u$ is more predisposed to telling the truth.
    \item The hierarchical position of $u$ ($L_u$) relative to the position of $v$ ($L_v$).
    %Here, the hierarchical position refers to the actual hierarchy of red nodes. If $v$ is higher in the hierarchy than $u$, $u$ will tend to lie more about $v$ to protect him. Conversely, if $v$ is below $u$ in the hierarchy, he might give away $u$'s true color to protect himself. 
 \end{itemize}   

Suppose $u$ is a red node.  In all lying scenarios, the probability that $u$ lies about another node $v$ is given by the following equations:

Equation~\ref{eq:v_red}  determines the probability $u$ will lie about a red node. $\frac{L_v}{L_u}$ indicates how far above $v$ is in the hierarchy compared to $u$ and $1-H_u$ is probability that $u$ will lie. %If $(1-H_u)*\frac{L_v}{L_u}$ greater than 1, the equation will return 1 as the probability. 
\begin{equation}\label{eq:v_red}
    P(u \; lie \; v | v=Red)= \min\{(1-H_u)*\frac{L_v}{L_u},1\}
\end{equation}

Equation~\ref{eq:v_blue} defines the probability $u$ will lie about a blue node. This depends on $u$'s honesty and is calculated as $(1-H_u)$.
\begin{equation}\label{eq:v_blue}
    P(u \; lie \; v | v=Blue)= (1-H_u)
\end{equation}

Now suppose that $u$ is a blue node.  $u$ may know nothing about red nodes.  This depends largely on whether the blue nodes are part of the same organization as the red nodes, but are simply not of interest to the user (e.g., blue and red nodes are all terrorists, and red nodes represent a subset of interest), or if the blue nodes represent individuals who are not part of the same organization as the red nodes (e.g., the red nodes are terrorists in a sea of blue node civilians).
    \vspace{-.8cm}
    \begin{center}
    \begin{table}[!htbp]
        \caption{Noordin Top network hierarchy assignment}
        \label{table:Noordin hierarchy}
    \begin{tabular}{|c||c|c|}
        \hline
        Role &  Hierarchy score & No. of nodes\\ 
        \hline
                \hline
        Strategist &5 & 10\\
        Commander; Religious Leader & 4 & 23\\
        Trainer/instructor; Bomb maker; Facilitator; Propagandist; Recruiter & 3 & 33\\
        Bomber/fighter; Suicide Bomber; Courier; Recon/Surveillance & 2 & 33\\
        Unknown & 1 & 40\\
        \hline
    \end{tabular}
    \vspace{-1cm}
    \end{table}
    \end{center}

%%%%%%%%%%%%%%%%%%%%%%%%%%%%%%%%%%%%%%%%%%%%%%%%%%%%%%%%%%%%%%%%%%%%%%

\begin{itemize}
    \item Lying scenario-1 (LS1): Blue nodes know about red nodes.  Here, $P(u \; lie \; v| u=Blue, v=Red)$ is determined using equation ~\ref{eq:v_red}, since blue nodes know about red nodes and their hierarchy. Additionally, blue nodes may lie about other blue nodes. $P(u \; lie \; v| u=Blue, v=Blue)$ is thus calculated using equation ~\ref{eq:v_blue}.
    
    \item Lying scenario-2(LS2): Blue nodes don't know about red nodes.  Here, blue nodes will simply say that all their neighbors are blue.  Because of this $P(u \; lie \; v| u=Blue, v=Blue)=0$ and $P(u \; lie \; v| u=Blue, v=Red)=1$
    
     %\item Lying strategy-3(LS3) :Blue nodes knows about red nodes, and lie about their red neighbors. Blue nodes do not lie about blue neighbors.\\
    %In this strategy blue nodes only lie about their neighboring red nodes, but not about blue nodes. Hence the $P(u \; lie \; v| u=Blue, v=Blue)=0$ , and true color of blue nodes are given. But still they lie about other red nodes and $P(u \; lie \; v| u=Blue, v=Red)$ is calculated using equation ~\ref{eq:v_red}. 
\end{itemize}

In all cases, if $P(u \; lie \; v)$ is greater than 1, it is rounded down to 1.

%In all lying strategies, if $u$ is a red node he will lie about red neighbors following equation~\ref{eq:v_red}. If $u$ is red and $v$ is a blue node, $u$ will lie about $v$ to protect other red neighbors based on honesty of $u$, following equation~\ref{eq:v_blue}. 

\subsection{Experiments}

Since our lying scenarios are probabilistic, the colors that nodes say about neighbors can change from one run of the algorithm to another. Additionally, the honesty assignment to a node also can change from one run to another.  Thus, for each network and lying scenario, we perform 25 runs of each monitor placement strategy.

In each run, we begin with a randomly selected red node.  For a fair comparison across different monitor placement strategies, we make sure that we run each monitor placement strategy with the same sets of starting nodes and honesty assignments.

In these experiments, the honesty of a node is drawn from a normal distribution, $h \sim \mathcal{N} (0.5,0.125)$.  In the Noordin Top network, ground truth hierarchy scores are known, and shown in Table~\ref{table:Noordin hierarchy}.  In the PokeC network, we set the hierarchy score to be the degree of the node. Given a particular lying scenario, a monitored node $u$ lies about a neighbor $v$'s color with probability $P(u \; lie \; v)$ as given in Section~\ref{subsection:lying-strategy}.

In each run, we consider budgets up to half the number of nodes in the network.  The Noordin Top network is small, and so we retrain {\sc RedLearn} after each monitor is placed.  The PokeC network is larger, so for the sake of efficiency, we train the learning model once per every $20$ monitors placed.

%%%%%%%%%%%%%%%%%%%%%%%%%%%%%%%%%%%%%%%%%%%%%%%%%%%%%%%%%%%%%%%%%%%%%%    
%%%%%%%%%%%%%%%%%%%%%%%%%%%%%%%%%%%%%%%%%%%%%%%%%%%%%%%%%%%%%%%%%%%%%%%%%%%%%%%%%%%%%%
\section{Results and Analysis}

In this section, we compare the results based on the various monitor placement strategies, and examine how performance is affected by (1) the presence or absence of homophily, (2) the lying scenario used by the nodes, and (3) the monitor placement budget.  

\begin{figure}[!h]
    \centering
    Homophily is present (original network)
    \vspace{-1cm}  
    
    \begin{subfigure}[b]{0.45\textwidth}
        \includegraphics[width=\textwidth, height=5.5cm, clip=True, trim=0cm 4cm 0cm 0cm]{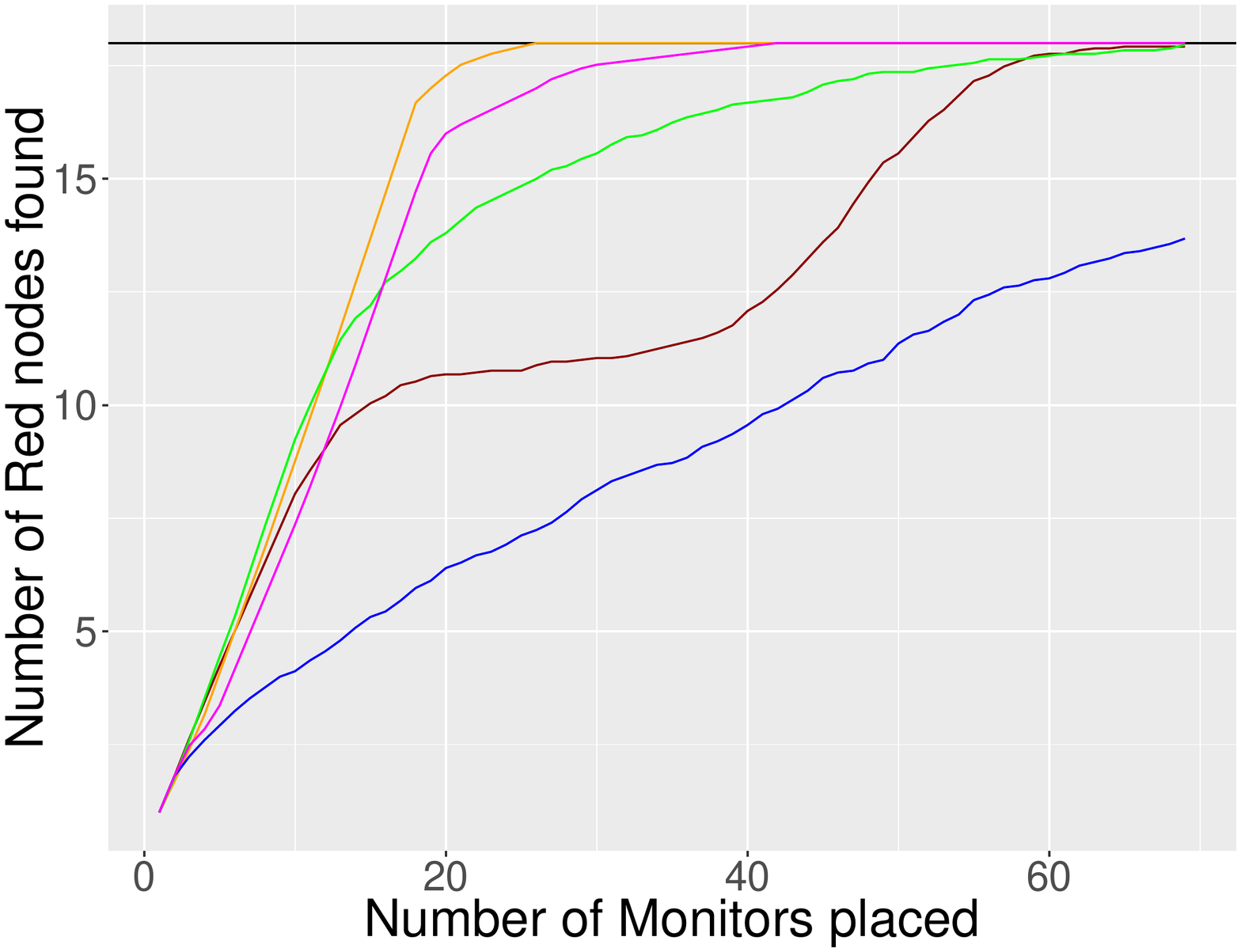}
        %\caption{LS1: All nodes lie about all colors }
        \label{fig:LS1 NoordinComs4}
    \end{subfigure}
    \quad
    \begin{subfigure}[b]{0.45\textwidth}
        \includegraphics[width=\textwidth, height=5.5cm, clip=True, trim=0cm 4cm 0cm 0cm]{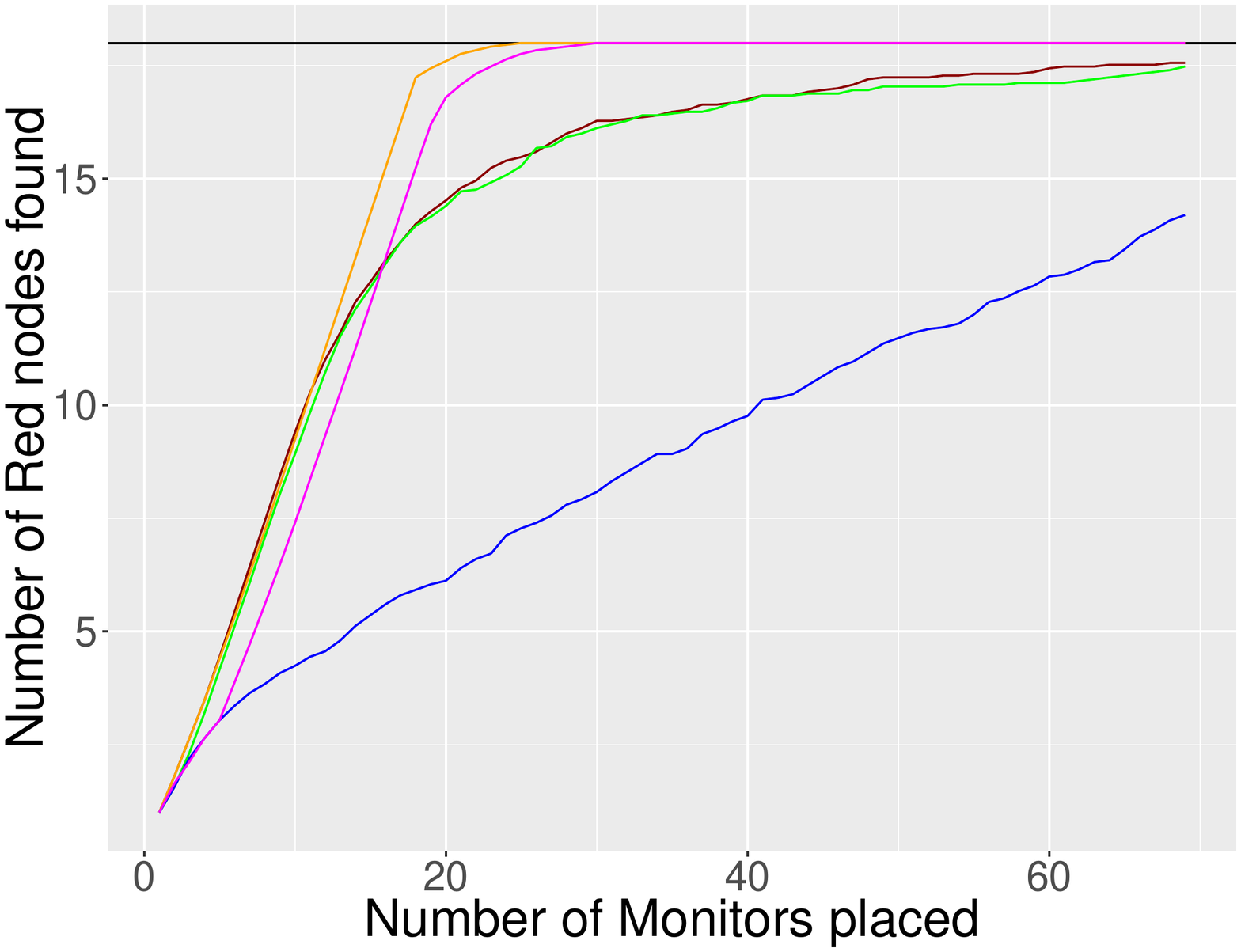}
        %\caption{LS2: Blue nodes say all nbrs. are blue}
        \label{fig:LS2 NoordinComs4}    
    \end{subfigure}

\vspace{-0.6cm}

      No Homophily (edges between red nodes are removed)
      
\vspace{-1cm}  

      \begin{subfigure}[b]{0.45\textwidth}
        \includegraphics[width=\textwidth, height=5.5cm, clip=True, trim=0cm 4cm 0cm 0cm]{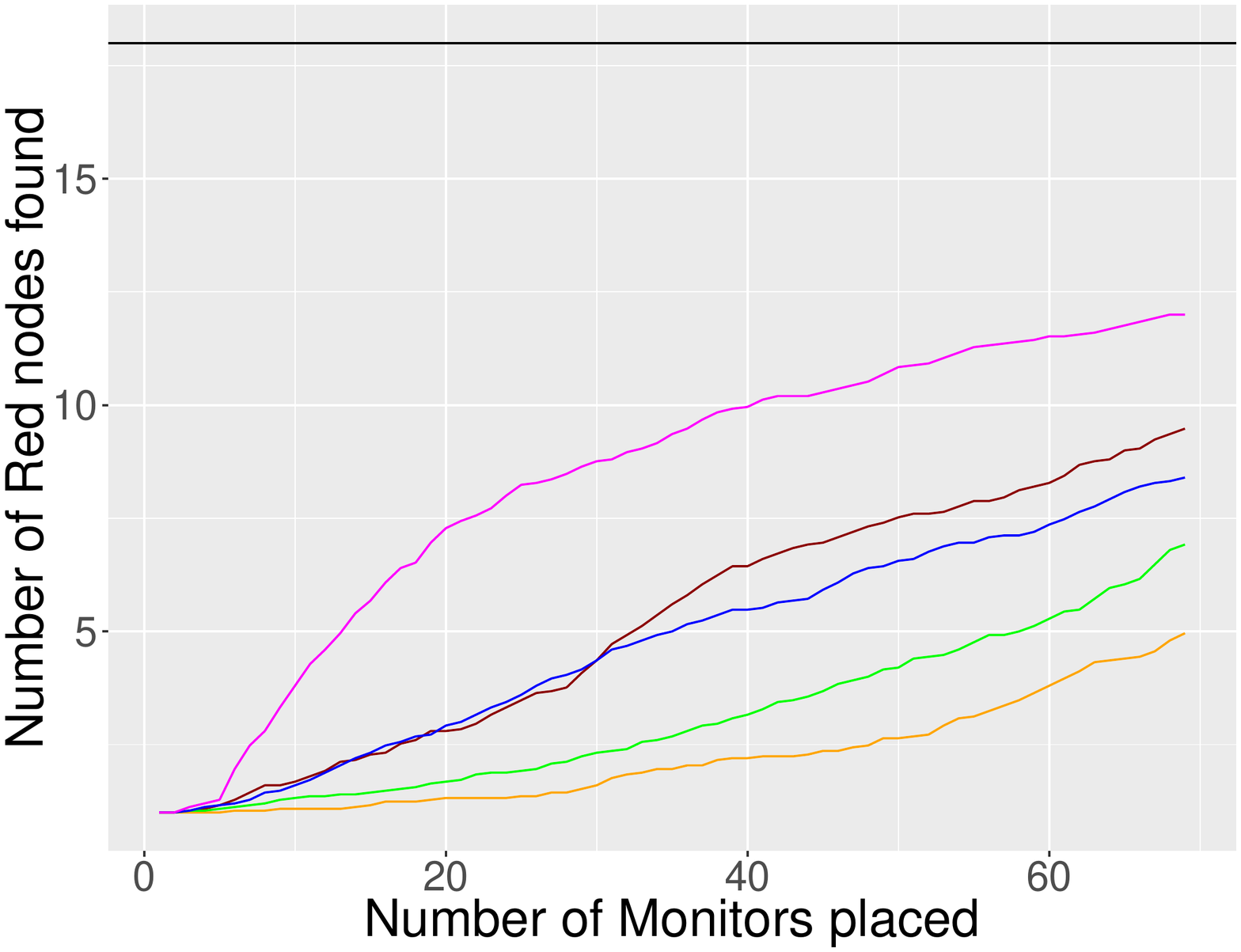}
        \caption{LS1: All nodes aware of red nodes.}
        \label{fig:LS1 NoordinComs4 no red}
    \end{subfigure}
    \quad
    \begin{subfigure}[b]{0.45\textwidth}
        \includegraphics[width=\textwidth, height=5.5cm,clip=True, trim=0cm 4cm 0cm 0cm]{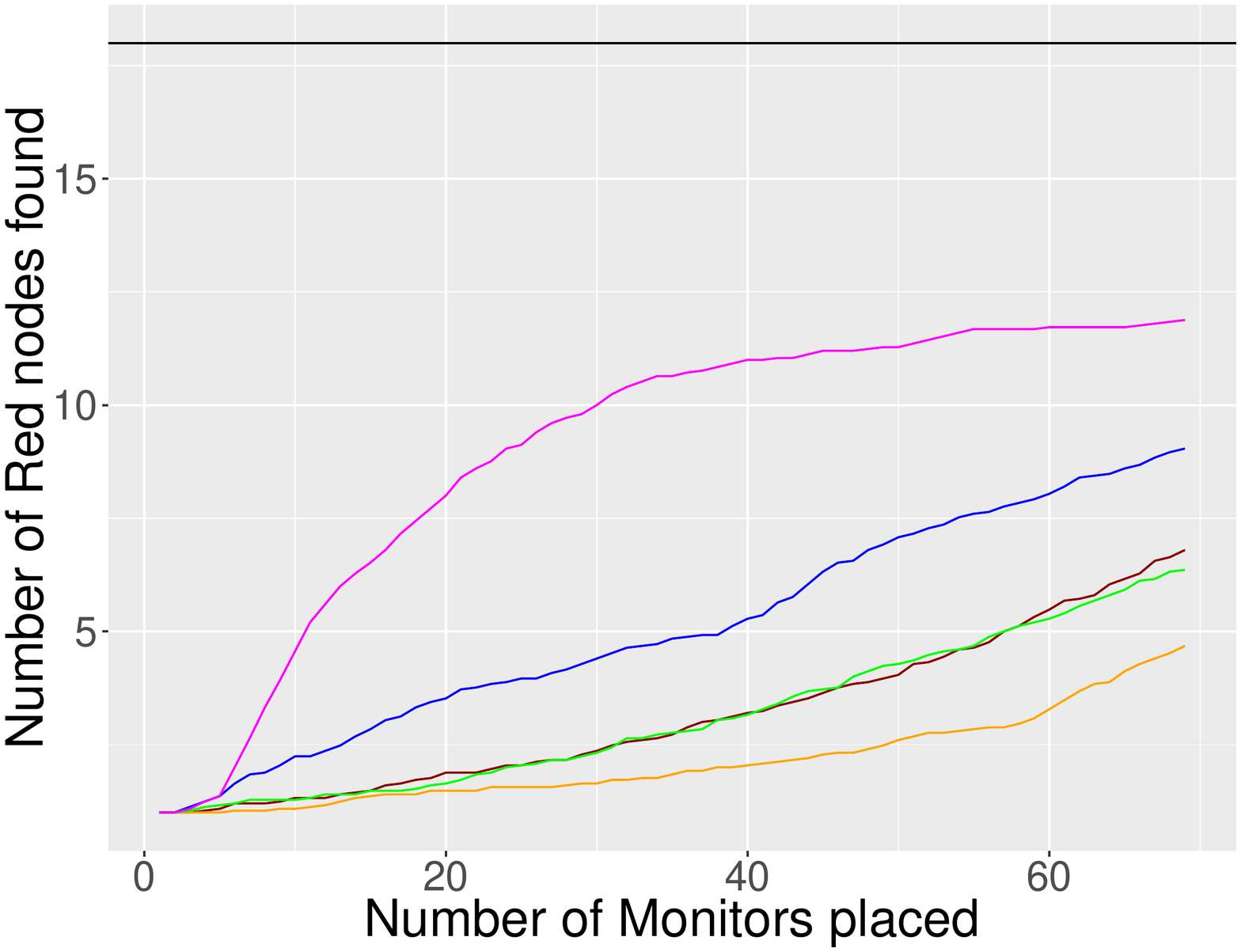}
        \caption{LS2: Only red nodes aware of red nodes.}
        \label{fig:LS2 NoordinComs4 no red}
    \end{subfigure}

\vspace{-.8cm} 
    \includegraphics[width=0.4\textwidth]{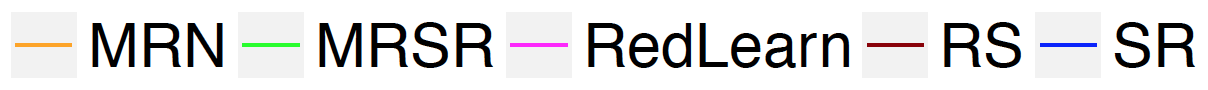}
    \vspace{1cm}
    \caption{Comparison of monitor placement strategies on the NoordinComs4 network.  The black line indicates the total number of red nodes present in the network.}
    \label{fig:2x2}
\end{figure}

As an example, Figure~\ref{fig:2x2}
%\ref{fig:NoordinComs4} and~\ref{fig:NoordinComs4_nored} 
shows results on the NoordinComs4 network with and without edges between red nodes, respectively.  When there is homophily, the problem becomes easy, and %(Figure~\ref{fig:NoordinComs4}), 
the simple strategy of monitoring the node with the most red neighbors (MRN) is best.  However, note that in both lying scenarios, {\sc RedLearn} is close behind the MRN strategy (because it needs time to train, it doesn't quite match the performance of MRN).

However, we see from the bottom figures that when edges between red nodes are removed, the MRN strategy performs very poorly.  
%Clearly, if red nodes have hidden their communications with other red nodes, then one can no longer simply rely on homophily.  
In this setting, {\sc RedLearn} performs much better than all comparison methods: it is able to learn the patterns and structural characteristics of red nodes, and by incorporating what neighbors say about a node, achieves strong performance.

Due to space constraints, we summarize results for the other cases in Tables~\ref{table:budget_homophily} and~\ref{table:budget_no_homophily}.  We see similar patterns across all networks: when there are edges between red nodes, it is enough to select the node with the most red neighbors; but when these edges are concealed, {\sc RedLearn} is the clear winner.

Even when there are edges between red nodes, {\sc RedLearn} usually achieves performance close to the MRN strategy.  There are some exceptions, such as the NoordinComs2 network. This typically occurs if there are very few red nodes: for instance, NoordinComs2, only 5 out of 139 nodes are red.  There is simply not enough information for {\sc RedLearn} to train on. 

This analysis shows that performance of the proposed algorithm does not rely on a particular network structure, or which lying scenario people might use. Therefore using learning based monitor placement is guaranteed to give better performance, especially in real networks, which are large.

\vspace{-1.3cm}
\begin{center}
\begin{table}[!h]
\caption{Comparison of the percentage of red nodes found from each monitor placement strategy.  Budgets include Low (10\% of the nodes), Medium (25\% of the nodes), and High (50\% of the nodes).  These networks exhibit homophily: edges between red nodes have not been removed.}
\label{table:budget_homophily}
\begin{subtable}{\textwidth}
\caption{Lying Scenario 1}
\scalebox{0.88}{
\begin{tabular}{c|c|c|c|c|c||c|c|c|c|c||c|c|c|c|c|}
\cline{2-16}
                                                                                                    & \multicolumn{5}{c||}{\textbf{Low Budget}}    & \multicolumn{5}{c||}{\textbf{Medium Budget}}    & \multicolumn{5}{c|}{\textbf{High Budget}}                   \\ \hline
\multicolumn{1}{|c|}{\textbf{\begin{tabular}[c]{@{}c@{}}Network/\\ Strategy\end{tabular}}} & RS & {\sc RdLrn} & MRN         & MRSR & SR & RS & {\sc RdLrn} & MRN         & MRSR & SR & RS          & {\sc RdLrn}       & MRN         & MRSR & SR \\ \hline
\multicolumn{1}{|c|}{NrdnComs1}                                                                & 28 & 74       & \textbf{97} & 52          & 32 & 43 & 97          & \textbf{100} & 77   & 51 & 97           & \textbf{100} & \textbf{100} & 92   & 75 \\ \hline
\multicolumn{1}{|c|}{NrdnComs2}                                                                & 42 & 37       & \textbf{62} & 48          & 42 & 61 & 72          & \textbf{100} & 72   & 55 & 99           & 93           & \textbf{100} & 91   & 81 \\ \hline
\multicolumn{1}{|c|}{NrdnComs3}                                                                & 33 & 63       & \textbf{83} & 52          & 27 & 59 & 89          & \textbf{100} & 77   & 46 & \textbf{100} & \textbf{100} & \textbf{100} & 97   & 73 \\ \hline
\multicolumn{1}{|c|}{NrdnComs4}                                                                & 54 & 60       & \textbf{70} & 66          & 28 & 63 & 98          & \textbf{100} & 90   & 48 & \textbf{100} & \textbf{100} & \textbf{100} & \textbf{100}  & 76 \\ \hline
\multicolumn{1}{|c|}{NrdnComs5}                                                                & 34 & 67       & \textbf{84} & 52          & 32 & 43 & \textbf{91} & \textbf{91}  & 75   & 46 & 88           & \textbf{91}  & \textbf{91}  & 86   & 67 \\ \hline
\multicolumn{1}{|c|}{PokeC age}                                                                     & 5  & 14       & \textbf{22} & 20          & 7  & 15 & 43          & \textbf{47}  & 39   & 21 & 48           & \textbf{73}  & 68           & 62   & 47 \\ \hline
\multicolumn{1}{|c|}{PokeC height}                                                                  & 14 & 14       & 21          & \textbf{23} & 11 & 36 & 32          & \textbf{48}  & 47   & 28 & \textbf{74}           & 64           & 73  & 69   & 54 \\ \hline
\end{tabular}
}
\label{table:low_budget}
\end{subtable}

\begin{subtable}{\textwidth}
\caption{Lying Scenario 2}
 \scalebox{0.88}{
\begin{tabular}{c|c|c|c|c|c||c|c|c|c|c||c|c|c|c|c|}
\cline{2-16}
                                                                                                    & \multicolumn{5}{c||}{\textbf{Low Budget}}    & \multicolumn{5}{c||}{\textbf{Medium Budget}}    & \multicolumn{5}{c|}{\textbf{High Budget}}                   \\ \hline
\multicolumn{1}{|c|}{\textbf{\begin{tabular}[c]{@{}c@{}}Network/\\ Strategy\end{tabular}}} & RS & {\sc RdLrn}       & MRN          & MRSR & SR & RS & {\sc RdLrn}        & MRN          & MRSR & SR & RS           & {\sc RdLrn}        & MRN          & MRSR & SR \\ \hline
\multicolumn{1}{|c|}{NrdnComs1} & 57          & 78    & \textbf{89} & 57   & 33 & 82 & \textbf{100} & \textbf{100} & 79   & 47 & 96 & \textbf{100} & \textbf{100} & 95   & 73 \\ \hline
\multicolumn{1}{|c|}{NrdnComs2} & 56          & 54    & \textbf{83} & 55   & 38 & 83 & 66           & \textbf{99}  & 82   & 52 & 94 & 89           & \textbf{100} & 94   & 74 \\ \hline
\multicolumn{1}{|c|}{NrdnComs3} & 50          & 70    & \textbf{76} & 52   & 34 & 77 & 85           & \textbf{100} & 80   & 50 & 99 & 97           & \textbf{100} & 99   & 77 \\ \hline
\multicolumn{1}{|c|}{NrdnComs4} & 68          & 62    & \textbf{74} & 67   & 28 & 92 & \textbf{100} & \textbf{100} & 91   & 50 & 98 & \textbf{100} & \textbf{100} & 97   & 79 \\ \hline
\multicolumn{1}{|c|}{NrdnComs5} & 59          & 64    & \textbf{88} & 59   & 32 & 79 & \textbf{91}  & \textbf{91}  & 79   & 50 & 89 & \textbf{91}  & \textbf{91}  & 90   & 74 \\ \hline
\multicolumn{1}{|c|}{PokeC age}      & 20          & 12    & \textbf{22} & 19   & 7  & 39 & 33           & \textbf{47}  & 39   & 21 & 62 & 60           & \textbf{68}  & 62   & 46 \\ \hline
\multicolumn{1}{|c|}{PokeC height}   & \textbf{23} & 12    & 21          & \textbf{23}   & 12 & 46 & 29           & \textbf{48}  & 46   & 28 & 69 & 62           & \textbf{73}  & 69   & 54 \\ \hline
\end{tabular}
}
\label{table:medium_budget}
\end{subtable}

\end{table}

       \vspace{-0.8cm}
\end{center}

%%%%%%%%%%%%%%%%%%%%%%%%%%%%%%%%%%%%%%%%%%%%%%%%%%%%%%%%%%%%%%%%%%%%%%
\begin{table}[!h]
\caption{Comparison of the percentage of red nodes found from each monitor placement strategy.  Budgets include Low (10\% of the nodes), Medium (25\% of the nodes), and High (50\% of the nodes).  These networks do not exhibit homophily among the red nodes: edges between red nodes have been removed.}
\label{table:budget_no_homophily}
\centering
\begin{subtable}{\textwidth}
\caption{Lying Scenario 1}
\scalebox{0.88}{
\begin{tabular}{c|c|c|c|c|c||c|c|c|c|c||c|c|c|c|c|}
\cline{2-16}
                                                                                                    & \multicolumn{5}{c||}{\textbf{Low Budget}}    & \multicolumn{5}{c||}{\textbf{Medium Budget}}    & \multicolumn{5}{c|}{\textbf{High Budget}}           \\ \hline
\multicolumn{1}{|c|}{\textbf{\begin{tabular}[c]{@{}c@{}}Network/\\ Strategy\end{tabular}}} & RS & {\sc RdLrn}       & MRN & MRSR & SR & RS & {\sc RdLrn}       & MRN & MRSR & SR & RS          & {\sc RdLrn}       & MRN & MRSR & SR \\ \hline
\multicolumn{1}{|c|}{NrdnComs1} & 16         & \textbf{33} & 12  & 14   & 22 & 38          & \textbf{46} & 20  & 27   & 36 & \textbf{72} & 64          & 40  & 48   & 58 \\ \hline
\multicolumn{1}{|c|}{NrdnComs2} & 30         & \textbf{70} & 21  & 26   & 35 & 50          & \textbf{82} & 26  & 41   & 55 & 86          & \textbf{94} & 52  & 63   & 84 \\ \hline
\multicolumn{1}{|c|}{NrdnComs3} & 21         & \textbf{59} & 12  & 14   & 22 & 57          & \textbf{82} & 16  & 28   & 44 & 76          & \textbf{99} & 41  & 57   & 76 \\ \hline
\multicolumn{1}{|c|}{NrdnComs4} & 12         & \textbf{30} & 6   & 8    & 12 & 31          & \textbf{52} & 11  & 15   & 28 & 53          & \textbf{67} & 28  & 38   & 47 \\ \hline
\multicolumn{1}{|c|}{NrdnComs5} & 13         & \textbf{35} & 10  & 11   & 16 & 30          & \textbf{51} & 12  & 18   & 26 & 52          & \textbf{55} & 28  & 34   & 40 \\ \hline
\multicolumn{1}{|c|}{PokeC age}      & 5          & \textbf{13} & 5   & 6    & 7  & 14          & \textbf{34} & 16  & 18   & 20 & 43          & \textbf{59} & 39  & 41   & 44 \\ \hline
\multicolumn{1}{|c|}{PokeC height}   & 13         & \textbf{14} & 5   & 7    & 11 & \textbf{33} & \textbf{33} & 15  & 19   & 27 & \textbf{69} & 59          & 37  & 44   & 52 \\ \hline
\end{tabular}
}
\label{table:low_budget_no_red}
\end{subtable}

\begin{subtable}{\textwidth}
\caption{Lying Scenario 2}
\scalebox{0.88}{
\begin{tabular}{c|c|c|c|c|c||c|c|c|c|c||c|c|c|c|c|}
\cline{2-16}
                                                                                                    & \multicolumn{5}{c||}{\textbf{Low Budget}}    & \multicolumn{5}{c||}{\textbf{Medium Budget}}    & \multicolumn{5}{c|}{\textbf{High Budget}}                   \\ \hline
\multicolumn{1}{|c|}{\textbf{\begin{tabular}[c]{@{}c@{}}Network/\\ Strategy\end{tabular}}} & RS & {\sc RdLrn}       & MRN & MRSR & SR & RS & {\sc RdLrn}       & MRN & MRSR & SR & RS          & {\sc RdLrn}       & MRN & MRSR & SR \\ \hline
\multicolumn{1}{|l|}{NrdnComs1} & 14         & \textbf{33} & 12  & 14   & 22 & 21            & \textbf{53} & 19  & 24   & 41 & 53          & \textbf{64} & 40  & 48   & 63 \\ \hline
\multicolumn{1}{|l|}{NrdnComs2} & 26         & \textbf{58} & 22  & 25   & 37 & 40            & \textbf{70} & 26  & 35   & 54 & 68          & 83 & 54  & 70   & \textbf{84} \\ \hline
\multicolumn{1}{|l|}{NrdnComs3} & 15         & \textbf{64} & 12  & 16   & 23 & 26            & \textbf{85} & 17  & 23   & 38 & 54          & \textbf{98} & 41  & 57   & 70 \\ \hline
\multicolumn{1}{|l|}{NrdnComs4} & 8          & \textbf{35} & 7   & 8    & 15 & 15            & \textbf{59} & 10  & 15   & 27 & 38          & \textbf{66} & 26  & 35   & 50 \\ \hline
\multicolumn{1}{|l|}{NrdnComs5} & 9          & \textbf{39} & 9   & 11   & 18 & 16            & \textbf{47} & 13  & 17   & 27 & 33          & \textbf{53} & 27  & 35   & 42 \\ \hline
\multicolumn{1}{|l|}{PokeC age}      & 6          & \textbf{10} & 5   & 6    & 7  & 16            & \textbf{26} & 14  & 16   & 18 & 42          & \textbf{59} & 39  & 41   & 44 \\ \hline
\multicolumn{1}{|l|}{PokeC height}   & 6          & \textbf{12} & 5   & 7    & 10 & 19            & \textbf{28} & 15  & 19   & 26 & 43          & \textbf{58} & 37  & 43   & 52 \\ \hline
\end{tabular}
}
\label{table:medium_budget_no_red}
\end{subtable}

\end{table}

%%%%%%%%%%%%%%%%%%%%%%%%%%%%%%%%%%%%%%%%%%%%%%%%%%%%%%%%%%%%%%%%%%%%%%
\section{Conclusions and Further Directions}
%therefore use this strategy if this is the type of network you have
Members of dark networks conceal information by nature, and while deceptive and sparse, they are still structured. Based on these properties,  we created and analyzed the results of several methods of sampling the networks to identify POI (red nodes). We developed a variety of natural sampling methods, and tested them both on a small real terrorist network as well as a larger social network.

We then created {\sc RedLearn}, a learning-based method for locating People of Interest in dark networks.  {\sc RedLearn} uses features from the simpler methods and learns how to identify red nodes in networks.  We showed that {\sc RedLearn} outperforms the other methods in cases where one cannot rely on homophily to identify red nodes.  %The only cases when it did not come in as clear winner are the cases when the count of POI is very small, not giving RedLearn a chance to learn in order to improve.  However, even in these cases it performed really well.

In our future work, one interesting direction is to consider the dynamicity of the network (both on the edge and node rate of birth and retirement), as well as a more sophisticated model of the concealed nodes and relationships.

\bibliographystyle{plain}
\bibliography{ArXive}

\end{document}